\documentclass[twocolumn,prb,amssymb,showpacs]{revtex4}
\usepackage{graphicx}

\begin{document}

\title{Nonohmic conductivity as a probe of crossover from diffusion to hopping in two dimensions}

\author{G.~M.~Minkov}
\author{A.~A.~Sherstobitov}
\author{O.~E.~Rut}
\author{A.~V.~Germanenko}

\affiliation{Institute of Physics and Applied Mathematics, Ural
State University, 620083 Ekaterinburg, Russia}

\date{\today}

\begin{abstract}
We  show that  the study of  conductivity nonlinearity gives a
possibility to  determine the condition when the diffusion
conductivity changes to the hopping one with increasing disorder.
It is experimentally shown that the conductivity  of single
quantum well GaAs/InGaAs/GaAs heterostructures behaves like
diffusive one down to value of order $10^{-2}e^2/h$.
\end{abstract}

\maketitle

Two opposite models are used to describe the conductivity in two
dimensional (2D) systems at low temperature. The first one
considers an electron as free particle with well defined
quasimomentum, which sometime scatters by static scattering
potential. The quantum corrections to the conductivity (due to
interference and interaction) determine in the main the
temperature and magnetic field dependences of the conductivity. It
is clear this model is valid for $k_Fl>1$, where $k_F$ and $l$ are
Fermi quasimomentum and mean free path, respectively.

The second model treats electrons as localized particles which
rarely hope from one localized state to another one. All the
results for this case are obtained in the framework of percolation
theory which is valid when the probability of the transitions
between the states is very small so that the dispersion of the
probabilities is exponentially large and percolation theory
becomes applicable. This theory gives\cite{Mott,ES}

\begin{equation}
\rho(T)=\rho_0(n,T)\exp({T_0/T})^p,\; 0<p<1 \label{eq1}.
\end{equation}
where $n$ is electron density and $T_0$ is constant depending on
localization length. The value of $p$ depends on relationship
between $k_B T$ and Coulomb gap. This expression is valid when the
exponent $({T_0/T})^p$ is larger than $10$.\cite{ES} The value of
$\rho_0(n,T)$ depends on overlapping of the wave functions of the
localized states and hopping mechanism, but for all cases it is
larger than $h/e^2$. So, the expressions for hopping conductivity
are valid when conductivity is less than $10^{-4} e^2/h$.

Thus the conductivity of 2D systems has to be described by theory
of quantum corrections when $\sigma>e^2/h$ and by theory of hoping
conductivity when $\sigma<10^{-4}e^2/h$. Over wide conductivity
range  $e^2/h>\sigma > 10^{-4}e^2/h$ both approaches are not
applicable in the strike sense. However, it is commonly accepted
that the conductivity mechanism in this range is hopping and
practically all the experimental data are treated in framework of
the theory of hopping conductivity.

One of theoretical reasoning behind this conclusion is the well
known Landauer formula which shows that the conductivity of one
open channel is $e^2/h$. For the first sight the 2D random network
of the open 1D channels should have the conductivity $e^2/h$ also.
This is true without taking into account interference. However,
the closed paths arise in such network and the interference of the
waves propagated in the different channels has to lead to
decreasing of the conductivity value. Therefore it seems that the
value of conductivity $e^2/h$ is not lower limit for the diffusive
conductivity when the transport is by delocalized states.

Experimentally, the conclusion on the conductivity mechanism comes
from an analysis of the temperature dependence of the
conductivity. When $\sigma<e^2/h$, the experimental  dependences
$\rho(T)$ can be satisfactorily  described by Eq.~(\ref{eq1}) with
$0.3\lesssim p\lesssim 0.8$.\cite{wrh1,wrh2,wrh3,wrh4,wrh5,wrh6}
Relaying on such value of power $p$, practically all the authors
conclude that the variable range hopping regime takes place.

We emphasize that a number of unusual features is observed in such
a 2D hopping regime as compared with the well-studied 3D variable
range hopping conductivity. The are the following.
\begin{enumerate}
\item The less than unity value of $p$ is observed up to
relatively high temperature, $T\simeq 5-10 $~K. The transition to
the nearest-neighbor hopping regime ($\epsilon_2$-regime) or to
the conductivity by free carriers ($\epsilon_1$-regime) is not
observed usually.

\item The value of $\rho_0$ is close to $h/e^2$ practically in all
cases independent of the structure type and carrier density.

\item The Hall coefficient is temperature independent,  its value
is determined by the electron density, $n$, down to very low
conductivity values of order $10^{-2} e^2/h$:  $R_H=(en)^{-1}$.

\item The negative magnetoresistance is observed down to $\sigma
\simeq (0.1-0.01) e^2/h$ and its shape is very close to that
observed for high conductivity $\sigma\gg e^2/h$, when it is
caused by suppression of interference quantum correction.
\end{enumerate}

Another approach to interpret the conductivity in this
intermediate range $e^2/h>\sigma > 10^{-4}e^2/h$ was used in
Ref.~\onlinecite{ourWLSL}. The authors attempted to trace the
changes in electron transport with decreasing conductivity caused
by increasing disorder starting from $\sigma\gg e^2/h$. They
analyzed not only the temperature dependences  of the conductivity
but the magnetic field dependences of the conductivity tensor
components at low and high magnetic field. A surprising conclusion
has been arrived: all the transport effects are well described
within the framework of the quantum correction theory down to
conductivity significantly less than $e^2/h$.

In the present paper we demonstrate that the study of nonohmic
conductivity can elucidate the low temperature conductivity
mechanism when $\sigma<e^2/h$. An idea of experiment will be clear
after consideration of the nature of conductivity nonlinearity in
diffusive and hopping regimes.

{\it Diffusive regime}. At low temperature, changing of the
conductivity $\sigma=j/E$ with electric field and, thus, with
injected power $Q=jE$ originates in this case from electron
heating. The heating leads to growth of the conductivity via
lowering  of the quantum corrections magnitude which is determined
by electron temperature.\cite{foot}

For steady state the injected power $Q$ is equal to the energy
relaxation rate $P$ which depends on the electron and lattice
temperature, $T_e$ and $T_l$, respectively. This equilibrium
determines the value of $T_e$ for given value of $T_l$. When the
energy relaxation rate $P(T_e,T_l)$ is governed by the interaction
with phonons it is equal to the difference of two identical
functions, one of them depends on the electron temperature and
another one depends on the lattice temperature:\cite{Price}
\begin{equation}
P(T_e,T_l)=F(T_e)-F(T_l).
\end{equation}
From this equation it follows that  $\partial P(T_e,T_l)/\partial
T_e$ does not depend on $T_l$
\begin{equation}
   { \partial P(T_e,T_l) \over \partial T_e}={ \partial F(T_e) \over \partial
   T_e}.
\end{equation}

Experimentally, the derivative $\partial P(T_e,T_l)/\partial T_e$
can be straightforwardly obtained from the dependences $\sigma(T)$
measured at low injected power and $\sigma(Q)$ measured at fixed
lattice temperature. Differentiating these dependences we can find
the experimental value of relative nonlinearity introduced as
\begin{equation}
\eta=\left.{\partial\sigma(Q)/\partial Q \over
\partial\sigma(T)/\partial T}\right|_{\sigma(Q)=\sigma(T)},
\label{eqNL}
\end{equation}
which is exactly equal to $(\partial F(T_e)/\partial T_e)^{-1}$.
Indeed,
\begin{eqnarray}
\left. {\partial\sigma(Q)/\partial Q  \over
\partial\sigma(T)/\partial T}\right|_{\sigma(Q)=\sigma(T)}&=&
{({\partial\sigma/\partial T_e})( {\partial T_e/\partial Q}) \over
\partial\sigma/\partial T}\\ \nonumber &=&{\partial T_e \over \partial Q} =
\left[{\partial  F(T_e) \over \partial
T_e}\right]^{-1}.\label{eq4}
\end{eqnarray}

\begin{figure}
\includegraphics[width=\linewidth,clip=true]{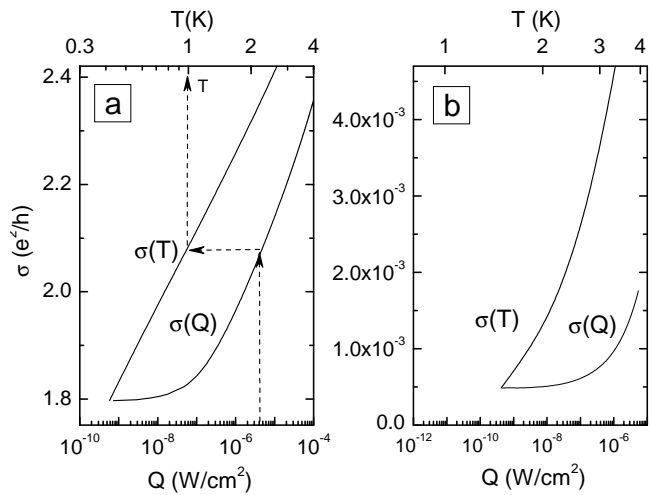}
\caption{Temperature and power dependences of the conductivity for
$\sigma(1.5\, K)=2.2~e^2/h$ (a)  and
$\sigma(1.5\,K)=6\times10^{-4}~e^2/h$ (b). }
 \label{f1}
\end{figure}

{\it Hopping regime}. First of all, the conductivity in this
regime depends both on the lattice and electron temperature.
Second, the changing of the conductivity with electric field
results not only from the electron heating but from the increasing
of the probabilities of hops and impact ionization of localized
states as well. Finally, the energy distribution function of the
electrons in electric field can deviate from the Fermi-Dirak
function and the approximation of electron temperature can fail.
All these effects have to lead to changing in relative
nonlinearity and, thus, the value of $\eta$ will depend on both
electron and lattice temperature.

Thus, we suggest a simple experimental way to distinguish the
diffusive conductivity from the hopping one. As long as the
conductivity remains diffusive the $T_e$-dependences of the
relative nonlinearity $\eta$ measured at different lattice
temperatures $T_l$ have to fall on common curve and this property
has to disappear when the conductivity becomes hopping.

Experimentally, we investigated the relative nonlinearity of the
conductivity for two types of the single quantum well
heterostructures GaAs/InGaAs/GaAs. The first one consists of 0.5
mkm-thick undoped GaAs epilayer, a Sn $\delta$-layer, a 9 nm
spacer of undoped GaAs, a 8 nm In$_{0.2}$Ga$_{0.8}$As well, a 9 nm
spacer of undoped GaAs, a Sn $\delta$-layer, and a 200 nm cap
layer of undoped GaAs.  The electron density and mobility were
$n=4\times 10^{15}$~m$^{-2}$ and $\mu=0.65$~m$^2$/(V~s),
respectively. The samples were mesa etched into Hall bars on which
basis field-effect transistors with an Ag or Al gate electrode
were fabricated. It was very important to use thick insulator
between gate electrode and 2D channel to decrease the influence of
voltage drop over the channel. We used the $10$~mkm organic
insulator.

\begin{figure}
\includegraphics[width=\linewidth,clip=true]{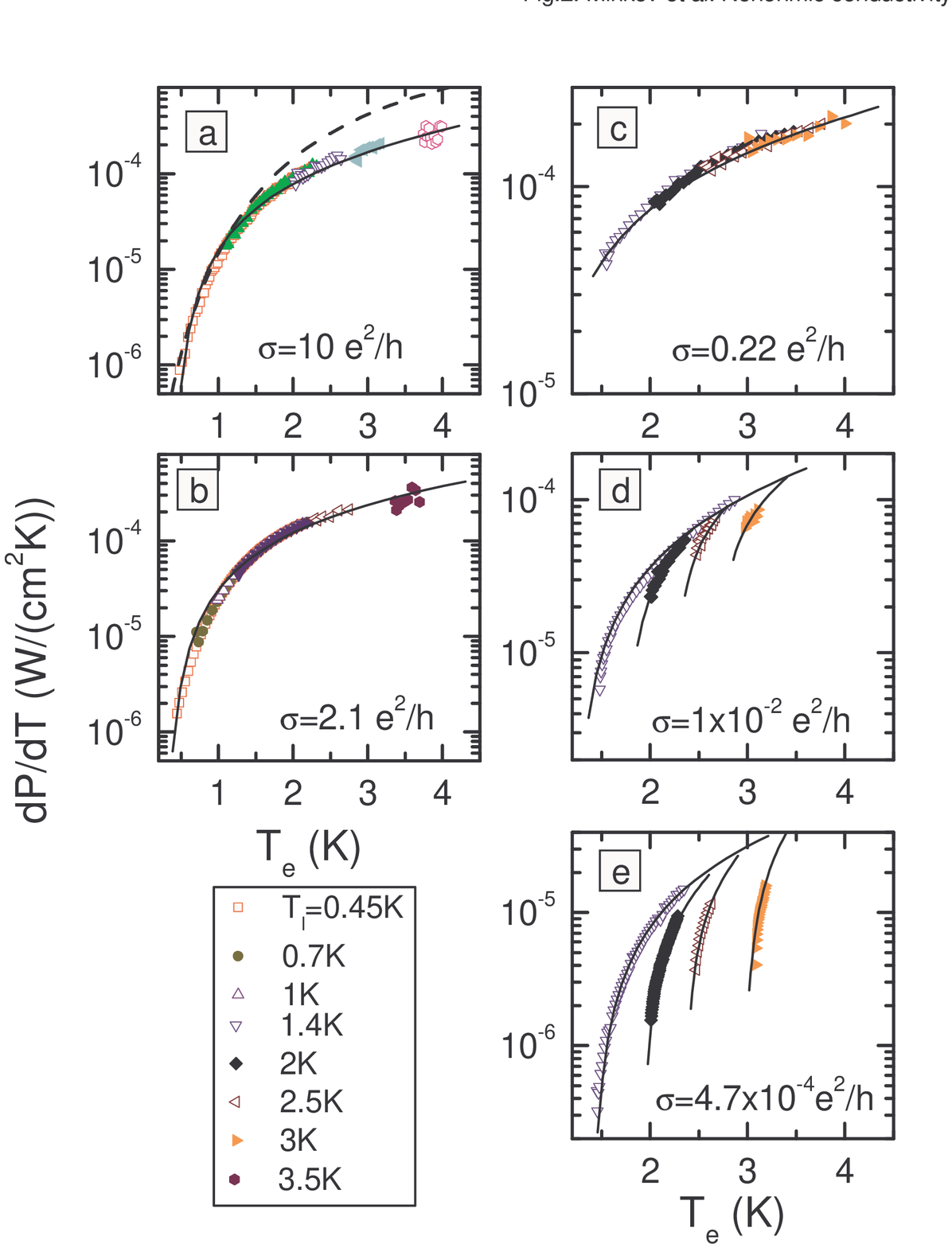}
\caption{The relative nonlinearity $\eta$ as a function of
electron temperature $T_e$ measured at different lattice
temperature $T_l$ for various conductivities. Dashed line in (a)
is calculated according to Ref.~\onlinecite{Price}. Solid lines
are provided as a guide for the eye.} \label{f2}
\end{figure}

Heterostructures of another type were analogous except that they
had no doping layers. The conductivity and electron density of
such samples varied by illumination from $ 10^{-4}$  to $10~e^2/h$
and from $1\times 10^{15}$ to $4\times 10^{15}$~m$^{-2}$,
respectively. Results obtained for both types of the structures
were close.

In what follows the results obtained for different electron
density will be referenced by the value of $\sigma$ measured at
$T=1.5$~K.

Figure~\ref{f1} shows the temperature dependence of the
conductivity measured in Ohmic regime and power dependences of
$\sigma$ for two significantly different conductivity values. The
dependences $\sigma(Q)$ were measured at sweeping of the current
through the sample from minus $j$ to plus $j$ during several
minutes. Within experimental error the $\sigma$-versus-$Q$ plots
were identical for both current directions.

At $\sigma> e^2/h$ the temperature dependence of the conductivity
is close to the logarithmic one  [Fig.~\ref{f1}(a)]. Such a
temperature dependence is determined by  the quantum corrections
and was discussed in detail in our previous
papers.\cite{ourWLSL,ourEE} The increase of the conductivity with
increasing injected power is result of electron heating and rise
of electron temperature from $T_l$ to $T_e$. For this case, the
electron temperature $T_e$ at given power $Q$ can be determined by
a standard manner\cite{standard} using the experimental
dependences $\sigma(T)$ measured at $Q\to 0$ and $\sigma(Q)$
measured at fixed temperature $T$ as shown in Fig.~\ref{f1}(a) by
arrows:  for each value of $Q$, $\sigma$ is used as a thermometer
giving $T_e$ as the temperature $T$ at which the same value of
$\sigma$ is measured for $Q\to 0$.

To obtain the relative nonlinearity we have numerically
differentiated  the experimental dependences $\sigma(T)$ and
$\sigma(Q)$ and found $\eta$ in accordance with Eq.~(\ref{eqNL}).
This quantity as a function of electron temperature taken at
different lattice temperature is presented in Fig.~\ref{f2}.

First we consider the results for highest conductivity $\sigma =
10~e^2/h$  which unambiguously corresponds to the diffusion regime
[Fig.~\ref{f2}(a)].  It is clearly seen  that $\eta$-versus-$T_e$
data obtained for different lattice temperatures fall on common
curve as it has to be in the diffusive regime. This curve is close
to theoretical dependence $\partial F(T_e)/\partial T_e$  when the
interaction with deformation and piezoelectric potential of the
acoustic phonons is taken into account.\cite{Price}

Such a data processing was carried out over the whole conductivity
range and the results are presented in Figs.~\ref{f2}(b)-(e). It
is evident that $\eta$-versus-$T_e$ data obtained for different
lattice temperatures fall on common curve down to $\sigma=
0.22~e^2/h$ and only at $\sigma = 1\times 10^{-2}e^2/h$ some
deviation arises. This deviation becomes drastic at $\sigma
=4.7\times 10^{-4}e^2/h$. It points to the fact that one or more
specific features of the hopping regime discussed above occur: the
conductivity becomes dependent both on the lattice and electron
temperature; the electric field leads to the conductivity change
not only via electron heating but through the increase of the
probability of hops and impact ionization; the approximation of
electron temperature fails. In means also that the way of finding
of the electron temperature  can be incorrect at $\sigma< 10^{-2}
e^/h$, and the quantity $T_e$ in Fig.~\ref{f2} is some effective
temperature.

In summary, we have proposed the way how studying the conductivity
nonlinearity one can determine the condition when the diffusion
regime changes to the hopping one. We have shown that for single
quantum well GaAs/InGaAs/GaAs heterostructures the conductivity
behaves like diffusive one down to $\sigma\approx (2-3)\times
10^{-2}e^2/h$. The conclusion that the low-temperature
conductivity of the disordered 2D systems remains diffusive down
to $\sigma\ll e^2/h$ agrees with that obtained from the studies of
the quantum corrections  to the conductivity at decreasing of
$k_Fl$ carried out in Ref.~\onlinecite{ourWLSL}.

This work was supported in part by the RFBR through Grants
No.~01-02-17003 and No.~03-02-16150, the INTAS through Grant No.
1B290, the Program {\it University of Russia}, the CRDF through
Grant No.~REC-005, and the Russian Ministry of Education through
Grant No.~A03-2.9-521.


\begin{thebibliography}{}


\bibitem{Mott} N.~F.~Mott and W.~D.~Twose., Adv. Phys. {\bf 10} 107
(1961).

\bibitem{ES} B.~I.~Shklovski and A.~L.~Efros, Electronic Properties of
DopedSemiconductors, Springer, Berlin, 1984.

\bibitem{wrh1}
H.W. Jiang, C.E. Jonson, K.L. Wang. Phys. Rev. B {\bf 46}, 12830
(1992).

\bibitem{wrh2}
H.W. Jiang, C.E. Jonson, K.L. Wang, S.T Hannahs. Phys. Rev. Lett.
{\bf 71}, 1439  (1993).

\bibitem{wrh3}
T. Wang, K.P. Clark, G.F. Spenser, A.M. Mack, W.P. Kirk. Phys.
Rev. Lett. {\bf 72}, 709 (1994).


\bibitem{wrh4}
F.V. Van Keuls, X.L. Hu, H.W. Jiang, A.J. Dahm Phys. Rev. B {\bf
56}, 1161 (1997).


\bibitem{wrh5}
C.H. Lee, Y.H. Chang, Y.W. Suen, H.H. Lin. Phys. Rev. B {\bf  58},
10629 (1998).

\bibitem{wrh6}
M. E. Gershenson, Yu. B. Khavin, D. Reuter, P. Schafmeister, and
A. D. Wieck. Phys. Rev. Lett. {\bf 85}, 1718 (2000).



\bibitem{ourWLSL}
G.~M.~Minkov, O.~E.~Rut, A.~V.~Germanenko, A.~A.~Sherstobitov,
B.~N.~Zvonkov, E.~A.~Uskova, and A.~A.~Birukov, Phys. Rev. B {\bf
65}, 235322 (2002).


\bibitem{foot} Estimations show
that approximation of electron temperature is valid at $n>
10^{14}$~m$^{-2}$.\cite{Dif}

\bibitem{Dif}
I.~L.~Drichko, A.~M.~Dyakonov, V.~D.~Kagan, A.~M.~Kreshchuk,
T.~A.~Polyanskaya, I.~G.~Savel'ev, I.~Yu.~Smirnov, and
A.~V.~Suslov. Fiz. Tekn. Poluprov. {\bf 31}, 1357 (1997) [
Semicond. {\bf 31} 1170 (1997)].

\bibitem{Price}
P.~J.~Price, J.Appl. Phys. {\bf 53}, 6863 (1982).

\bibitem{ourEE}
G.~M.~Minkov, O.~E.~Rut, A.~V.~Germanenko, A.~A.~Sherstobitov,
V.~I.~Shashkin, O.~I.~Khrykin, and B.~N.~Zvonkov, Phys. Rev. B
{\bf 67}, 205306 (2003).

\bibitem{standard}
R.~Fletcher, Y.~Feng, C.~T.~Foxon, and J.~J.~Harris. Phys. Rev. B
{\bf 61}, 2028 (2000).



\end{thebibliography}
\end{document}